\documentclass[preprint,11pt,showkeys,showpacs,nofootinbib,floatfix]{revtex4}
\usepackage{amssymb}
\usepackage{amsfonts}
\usepackage{amsmath}
\usepackage{graphicx}
\usepackage{subfigure}
\usepackage[dvipdfm,
            pdfstartview=FitH,
            bookmarksnumbered=true,
            bookmarksopen=true,
            colorlinks,
            pdfborder=001,
            linkcolor=green,
            anchorcolor=green,
            citecolor=red
            ]{hyperref}
\usepackage[toc,page,title,titletoc,header]{appendix}
\begin{document}
\title{Magnetic-field effects on $p$-wave phase transition in Gauss-Bonnet gravity}
\author{Ya-Bo Wu$^1$}
\thanks{E-mail address:ybwu61@163.com}
\author{Jun-Wang Lu$^1$}
\author{Yong-Yi Jin$^2$}
\author{Jian-Bo Lu$^1$}
\author{Xue Zhang$^1$}
\author{Si-Yu Wu$^1$}
\author{Cui Wang$^1$}
\affiliation{
$^1$Department of Physics, Liaoning Normal University, Dalian 116029, P.~R.~China\\
$^2$China Criminal Police University, Shenyang, P.~R.~China}
\begin{abstract}
In the probe limit, we study the holographic $p$-wave phase transition in the Gauss-Bonnet gravity via numerical and analytical methods. Concretely, we study the influences of the external magnetic field on the Maxwell complex vector model in the five-dimensional Gauss-Bonnet-AdS black hole and  soliton backgrounds, respectively. For the two backgrounds, the results show that the magnetic field enhances the superconductor phase transition in the case of the lowest Landau level, while the increasing Gauss-Bonnet parameter always hinders the vector condensate. Moreover, the Maxwell complex vector model is a generalization of the SU(2) Yang-Mills model all the time. In addition, the analytical results backup the numerical results.  Furthermore, this model might provide a holographic realization for the  QCD vacuum instability.
\end{abstract}
\maketitle
\section{Introduction}
The gauge/gravity duality~\cite{Maldacena1998,Gubser105} can map a strongly coupled conformal field theory onto a weak gravity system, which hence provides us a new method to study the condensed matter physics involving the strong interaction, especially the high temperature superconductors.

The simplest superconductor model was constructed numerically in the four-dimensional Schwarzschild anti-de Sitter(AdS) black hole coupled to a  Maxwell complex scalar field~\cite{Hartnoll2008}. When the temperature falls over a critical value, the AdS solution becomes instable to producing a scalar ``hair" that spontaneously breaks the U(1) symmetry of the system, which therefore models the $s$-wave superconductor phase transition, whereafter, holographic $p$-wave and $d$-wave superconductors were built, respectively, for details, see Refs.~\cite{Gubser2008a,Chen:2010mk}. Meanwhile, the critical behavior of the holographic model was studied analytically via the Sturm-Liouville (SL) eigenvalue method in Ref.~\cite{Siopsis}. However, all above works were investigated in the probe limit, where the backreaction of the matter field on the background is neglected, beyond which the holographic model was further investigated in Ref.~\cite{Hartnoll:2008kx}.  In addition to above conductor/superconductor models, the insulator/superconductor phase transition was modeled in the AdS soliton background~\cite{Nishioka:2009zj}. Moreover, the holographic superconductor models were studied in the system including the magnetic field~\cite{Albash:2008eh,Nakano:2008xc,Cai:2011tm,Chernodubjia}, and were also extended in various backgrounds, especially in the Gauss-Bonnet spacetimes and the Lifshitz spacetimes, for example, in Refs.~\cite{Gregory:2009fj,Cai:2010cv,Li:2011xja,Cai:2010zm,Pan:2009xa, Gangopadhyay:2012np,Pan:2011ah,Buyanyan,Lu:2013tza,Zhao:2013pva,Momeni:2012tw,Momeni:2012uc}, where the results showed that the increasing magnetic field and the Gauss-Bonnet parameter as well as the Lifshitz exponent hinder the phase transition.

On the other hand, the authors of Ref.~\cite{Cai:2013pda} proposed a holographic $p$-wave superconductor model in the four-dimensional Schwarzschild AdS black hole coupled to a Maxwell complex vector (MCV) field in the probe limit. The results showed that, for the lowest Landau level, even without the charge density, the vector condensate can be triggered when the applied magnetic field increases to a critical value, which is reminiscent of the QCD vacuum phase transition~\cite{Wong:2013rda, Chernodub:2010qx,Chernodub:2011mc}, while for the excited Landau level (i.e., the excited Landau level), the strong magnetic field protects the stability, which is similar to the case of the ordinary superconductors~\cite{Albash:2008eh,Nakano:2008xc,Zhao:2013pva}. In Ref.~\cite{Cai:2013kaa}, the holographic insulator/superconductor phase transition induced by the magnetic field was studied in the five-dimensional AdS soliton coupled to such a MCV field and the SU(2) Yang-Mills (YM) field, respectively. It was shown that the results are similar to the case in the black hole, and the MCV model is a generalization of the SU(2) model with general mass, charge and the magnetic moment. Going away the probe limit, the superconductor model without the magnetic field was further studied in the black hole and soliton backgrounds, respectively, in Refs.~\cite{Cai:2013aca,Li:2013rhw,Cai:2014ija}, and the rich phase structures were exhibited, especially the ``retrograde condensate". For this MCV model, we studied the critical behavior induced by the magnetic field in the Lifshitz black hole~\cite{wuyb}. It was found that the increasing Lifshitz dynamical exponent enhances the vector condensate in the case of the lowest Landau level, and inhibits the phase transition with the excited Landau level. Hence, we wonder whether such interesting dependence of the critical behavior on the background and the magnetic field  still exists in the higher curvature theory, for example, the Gauss-Bonnet gravity, which is our motivation in this paper.

Based on the above motivation, following Refs.~\cite{Cai:2013pda,Cai:2013kaa}, we will study the holographic $p$-wave phase transition induced by the applied magnetic field in the Gauss-Bonnet-AdS backgrounds coupled to the MCV field at the probe approximation. For both the Gauss-Bonnet-AdS black hole and soliton backgrounds, the results show that the increasing Gauss-Bonnet parameter $\alpha$ hinders the phase transition whether the Landau level is the lowest or not. Moreover, the magnetic field enhances the vector condensate for the lowest Landau level but hinders the phase transition for the excited Landau level. In addition, the MCV model is always a generalization of the SU(2) YM model. Furthermore, the analytical results agree with the numerical results.

This paper is organized as follows. In Sec.~II, in the probe limit, we study the holographic $p$-wave superconductor phase transition induced by the applied magnetic field  in the Gauss-Bonnet-AdS black hole background coupled to the MCV field and the SU(2) field, respectively. Both the MCV field and the SU(2) field are studied in Gauss-Bonnet-AdS soliton background in Sec.~III. The final section is devoted to conclusions and  discussions.

\section{$p$-wave phase transition in Gauss-Bonnet-AdS  black Hole}
In this section, we study the holographic $p$-wave superconductor phase transition induced by the magnetic field in the Gauss-Bonnet-AdS black hole coupled to the MCV field. To see the relation between the MCV model  and the SU(2) gauge field model, we also deduce the equations of motion for the SU(2) model in the presence of the applied magnetic field.

The five-dimensional Ricci flat Gauss-Bonnet-AdS black hole reads~\cite{Cai:2001dz}
\begin{eqnarray}\label{Gmetric}
ds^2&=&-r^2f(r)dt^2+\frac{dr^2}{r^2f(r)}+r^2(dx^2+dy^2+dz^2),\\
f(r)&=&\frac{1}{2\alpha}\left(1-\sqrt{1-\frac{4\alpha}{L^2}(1-\frac{ML^2}{r^4})}\right),\nonumber
\end{eqnarray}
where $M$ and $L$ denote the mass of the black hole and the AdS radius, respectively, while the constant $\alpha$ stands for the Gauss-Bonnet coupling, which has the upper bound, i.e., the so-called Chern-Simons limit $\alpha=L^2/4$. However, if we consider further the constraints of the causality via the gauge/gravity duality, for example, in Ref.~\cite{Buchel:2009tt}, the Gauss-Bonnet parameter obeys the range $-7L^2/36\leq\alpha\leq9L^2/100$, which we will work in this paper.  The Hawking temperature of the black hole is given by $T=\frac{r_+}{\pi L^2}$, where $r_+=\sqrt[4]{ML^2}$ denotes the location of the horizon satisfying $f(r)=0$. Near the asymptotical infinity, the metric function has the form
\begin{equation}\label{asyform}
f(r)\sim\frac{1}{2\alpha}\left(1-\sqrt{1-\frac{4\alpha}{L^2}}\right).
\end{equation}
For simplicity, we define an effective AdS radius as
\begin{equation}\label{Adsr}
L^2_{eff}=\frac{2\alpha}{1-\sqrt{1-\frac{4\alpha}{L^2}}}.
\end{equation}
\subsection{Maxwell complex vector model}
As Ref.~\cite{Cai:2013pda}, we take the action of matter field  including a Maxwell field and a complex vector field
\begin{eqnarray}
\mathcal{S}_{MCV}&=&\frac{1}{16\pi G_5}\int dx^5\sqrt{-g}\Big{(}-\frac{1}{4}F_{\mu\nu}F^{\mu\nu}-\frac{1}{2}(D_\mu\rho_\nu-D_\nu\rho_\mu)^\dag(D^\mu\rho^\nu-
D^\nu\rho^\mu)\nonumber\\
&&-m^2\rho^\dag_\mu\rho^\mu+iq\gamma\rho_\mu\rho^\dag_\nu F^{\mu\nu}\Big{)}\label{Lvector},
\end{eqnarray}
 where $F_{\mu\nu}$ stands for the strength of the U(1) gauge field $A_\mu$, and the operator ``$D_\mu$" denotes the covariant derivative $D_\mu=\nabla_\mu-iq A_\mu$, while $m$ ($q$) is the mass (charge) of the vector field $\rho_\mu$. It is worth noting that the constant $\gamma$ not only characterizes the strength of interaction between the vector field $\rho_\mu$ and the gauge field but also is regarded as the effective gyromagnetic ratio of $\rho_\mu$.

Varying the action (\ref{Lvector})  with respect to $\rho_\mu$ and $A_\mu$, respectively, equations of motion in term of $\rho_\mu$ and  $A_\mu$ read
\begin{eqnarray}
 D^\nu(D_\nu\rho_\mu-D_\mu\rho_\nu)-m^2\rho_\mu+iq\gamma\rho^\nu F_{\nu\mu}&=&0,\label{EOMrho} \\
  \nabla^\nu F_{\nu\mu}-iq(\rho^\nu(D_\nu\rho_\mu-D_\mu\rho_\nu)^\dag-\rho^{\nu\dag}(D_\nu\rho_\mu-D_\mu\rho_\nu))
  +iq\gamma\nabla^\nu(\rho_\nu\rho^\dag_\mu -\rho^\dag_\nu\rho_\mu)&=&0.\label{EOMphi}
\end{eqnarray}
Comparing with the action of the Gauss-Bonnet-AdS black hole background, in this paper, we regard the matter sector (\ref{Lvector})  as a perturbation. This is the so-called probe limit, where the Einstein equations of motion decouple from Eqs.~(\ref{EOMrho}) and (\ref{EOMphi}). In spite of this, it is believed that the main physics can still be revealed at this approximation.

In order to construct the vector condensate triggered by the applied magnetic field, it is turned out that we take the following ansatzs for the vector field $\rho_\mu$ and  the gauge field $A_\mu$ as
\begin{eqnarray}
\rho_\nu dx^\nu&=&\epsilon\rho_x(r,x)e^{ipy}dx+\epsilon\rho_y(r,x)e^{ipy}e^{i\theta}dy,\label{MCVansatz}\\
A_\nu dx^\nu&=&\phi(r) dt+B x dy,\label{MCVansAu}
\end{eqnarray}
where $\epsilon$ represents a small parameter characterizing the deviation from the critical point between the normal phase and the condensed phase, and $\theta$ is related to the phase difference between the $x$ and $y$ components of the vector field $\rho_\mu$, while $p$ is a constant that is used to construct the vortex lattice solution. Without loss of generality, we choose  $\rho_x(r,x),~\rho_y(r,x)$ and $\phi(r)$ as real functions.

Substituting the above ansatzs (\ref{MCVansatz}) and (\ref{MCVansAu}) into Eq.~(\ref{EOMrho}), we can deduce the equations for $\rho_x$ and $\rho_y$ at linear order. However, to satisfy the equations of motion of this vector model, the phase difference $\theta$ can only be chosen as $\theta_\pm=\pm\frac{\pi}{2}+2 n\pi$, where $n$ denotes an arbitrary integer~\cite{Cai:2013pda}. Making further a variable separation as $\rho_x(r,x)=\psi_x(r)U(x)$ and $\rho_y(r,x)=\psi_y(r)V(x)$, we can obtain the following equations:
\begin{eqnarray}
\psi_x (r) \dot{U}(x)\pm ( q B x-p)\psi_y (r) V(x)=0,\label{eomX} \\
\psi'_x (r) \dot{U}(x)\pm (q B x-p)\psi'_y (r) V(x)=0,\label{eomY}
\end{eqnarray}
\begin{eqnarray}
\psi''_x+\left(\frac{3}{r}+\frac{f'}{f}\right)\psi'_x-\frac{m^2}{r^2f}\psi_x+\frac{q^2\phi^2}{r^4f^2}\psi_x+\ \ \ \ \ \ \ \ \ \ \ \ \ \ \ \ \ \ \ \ \ \ \ \ \ \ & &\nonumber \\
\ \ \ \ \ \ \ \ \ \ \ \frac{\psi_x}{r^4f}\left(-(q B x-p)^2\pm \left((p-q B x)\frac{\dot{V}}{U}+\gamma q B\frac{ V}{U}\right)\frac{\psi_y}{\psi_x}\right)&=&0, \label{eomvarpx}\\
\psi''_y+\left(\frac{3}{r}+\frac{f'}{f}\right)\psi'_y-\frac{m^2}{r^2f}\psi_y+\frac{q^2\phi^2}{r^4f^2}\varphi_y
+\ \ \ \ \ \ \ \ \ \ \ \ \ \ \ \ \ \ \ \ \ \ \ \ \ \ \ &&\nonumber\\
\frac{\psi_y}{r^4f}\left(\frac{\ddot{V}}{V}\pm \left((1+\gamma)q B \frac{U}{V}+ ( q B x-p )\frac{\dot{U}}{V}\right)\frac{\psi_x}{\psi_y}\right)&=&0,\label{eomvarpy}
\end{eqnarray}
where the prime (dot) represents the derivative with respect to $r$ ($x$), while the sign ``$\pm$" corresponds to $\theta_\pm$. By introducing a real constant $c$, we can write out the consistent forms of Eqs.~(\ref{eomX}) and (\ref{eomY}) as
\begin{equation}\label{eomC}
\varphi_x=c\varphi_y, \ \ \ \dot{U}\pm \frac{(qBx-p)}{c} V=0,
\end{equation}
 from which we can separate Eqs.~(\ref{eomvarpx}) and (\ref{eomvarpy}) into three equations:
\begin{eqnarray}
\psi''_x+\left(\frac{3}{r}+\frac{f'}{f}\right)\psi'_x-\frac{m^2}{r^2f}\psi_x+\frac{q^2\phi^2}{r^4f^2}
\psi_x-\frac{E}{r^4f}\psi_x&=&0,\label{eigvarpx} \\
\ddot{U}\pm\frac{ q B(1+\gamma)}{c}V-(q B x-p)^2U+ E U&=&0,\label{eigX}\\
\ddot{V}\pm c q B(1+\gamma)U-(q B x-p)^2V +E V&=&0, \label{eigY}
\end{eqnarray}
with the eigenvalue $E$. By replacing $c\rightarrow \frac{1}{c}$, it is evident that Eqs.~(\ref{eigX}) and (\ref{eigY}) are the same as the ones in Refs.~\cite{Cai:2013pda,Cai:2013kaa,wuyb}. For simplicity, we choose $c^2=1$ in order to give exact solutions for $U(x)$ and $V(x)$. Defining further a new function  $X(x)=U(x)-V(x)$ and a variable $\xi=\sqrt{|q B|}(x-\frac{p}{q B})$, we can obtain the harmonic-oscillator equation about $X$ by subtracting  Eq.~(\ref{eigX}) from Eq.~(\ref{eigY}) as
\begin{equation}\label{hermi}
\frac{d X^2(\xi)}{d \xi^2}+(\lambda-\xi^2)X(\xi)=0,
\end{equation}
 where the constant $\lambda=\frac{E\mp q c B(1+\gamma)}{|q B|}$. The regular and bounded solution to Eq.~(\ref{hermi}) can be given in terms of the Hermite function $H_l$
\begin{equation}\label{solupsi}
X(\xi)=N_l e^{-\frac{\xi^2}{2}}H_l(\xi),
\end{equation}
where $N_l$ is a normalization constant, and the corresponding eigenvalue (i.e., the so-called Landau level) is of the form
\begin{equation}\label{Enphi}
E_l=(2l+1)|q B|\pm q c B(1+\gamma),
\end{equation}
where $l$ is a non-negative integer. Since the solution (\ref{solupsi}) is independent of the Gauss-Bonnet parameter $\alpha$, we can conclude that the vortex lattice solution is the same as the case in Refs.~\cite{Cai:2013pda,Cai:2013kaa ,wuyb} and has nothing to do with the black hole background, so we will not display the vortex lattice in this paper for brevity. Because of the presence of the nonminimal coupling $\gamma$ between the gauge field and the vector field $\rho_\mu$, the lowest Landau level (by choosing $l=0$, sign($qcB$)$=\mp$ and $\gamma>0$) has an interesting form as $E_0^L=-|\gamma q B|$, which is believed to give unusual effect on the critical point.

Introducing a dimensionless coordinate $u=\frac{r_+}{r}$, Eq.~(\ref{eigvarpx}) can be rewritten as
\begin{equation}\label{BHMCVpsiu}
\psi _x''(u)+\left(\frac{f'(u)}{f(u)}-\frac{1}{u}\right) \psi _x'(u)+ \left(\frac{q^2 \phi (u)^2}{r_+^2 f(u)^2}-\frac{E_l}{r_+^2 f(u)}-\frac{m^2}{u^2
   f(u)}\right)\psi _x(u)=0.
\end{equation}
 In the remainder of this paper, we will set $r_+=1$ and $L=1$ for the numerical calculation. Near the boundary $u\rightarrow0$, the general falloff of $\psi_x(u)$ is of the form
\begin{equation}\label{expvarx}
\psi_x(u)=\psi_{x-}u^{\Delta_-}+\psi_{x+}u^{\Delta_+},
\end{equation}
where $\Delta_\pm=1\pm\sqrt{1+ m^2 L_{eff}^2}$. According to the gauge/gravity dual dictionary, $\psi_{x-}$ and $\psi_{x+}$ correspond to the source and the vacuum expectation value of the dual vector operator $J_x$ in the boundary field theory.  To meet the requirement that the U(1) symmetry as well as the spatially rotational symmetry of the system are spontaneously broken, we impose the source-free condition, i.e., $\psi_{x-}=0$.

Plugging the ansatzs (\ref{MCVansatz}) and (\ref{MCVansAu}) into Eq.~(\ref{EOMphi}) gives the equation of motion for $\phi(r)$
\begin{equation}
\phi''(r)+\frac{3}{r}\phi'(r)=0,
\end{equation}
where we have only considered the perturbation up to the linear order.  In order to ensure the finite form of the gauge potential $A_\mu$ at the horizon, $\phi(r_+)$ is usually required to be vanishing, while near the boundary $r\rightarrow\infty$, the leading term (the coefficient of the subleading term) of the general falloff is regarded as the chemical potential $\mu$ (the charge density $\rho$) in the dual field theory. For convenience of the following calculation,  we write the solution of $\phi$ by the dimensionless coordinate
\begin{equation}
\phi(u)=\mu\left(1-u^2\right).
\end{equation}

 Using the shooting method, we now  study the influences of both the applied magnetic field and the Gauss-Bonnet parameter $\alpha$ on the critical point for this vector model with the lowest Landau level~ ($E^L_0=-|q\gamma B|$) by choosing $l=0$, sign($qcB$)$=\mp$, and $\gamma>0$. For simplicity, we first consider the case without the charge density (i.e., $\phi(u)=0$). Since the black hole solution becomes instable to developing a vector condensate near the critical point, we will encounter a marginally stable mode for Eq.~(\ref{BHMCVpsiu}). Therefore, for a given $m^2$ and  $\alpha$, there are only some special $\zeta$ that can satisfy the equation. For the sake of  numerical calculation, we further introduce a new function $\psi_x (u)=u^{\Delta_-}R(u)$, which yields the equation
\begin{equation}\label{eomzeta}
R''+ \left(\frac{f'}{f}+\frac{2 \Delta_--1}{u}\right)R'+\frac{ \Delta _- u f'+ \left(\Delta _--2\right) \Delta _- f- m^2}{\pi ^2 f u^2}R+\frac{\zeta}{\pi ^2 f}R=0,
\end{equation}
with $\zeta=|\gamma q B|/T^{2}$, where we have used the definition of the temperature $T=\frac{r_+}{\pi}$.   To solve this equation, in addition to the source-free condition at infinity, we also impose the regular condition at the horizon. Without loss of generality, we take $R(1)=1$.

\begin{figure}
\centering
\includegraphics[width=2.9in]{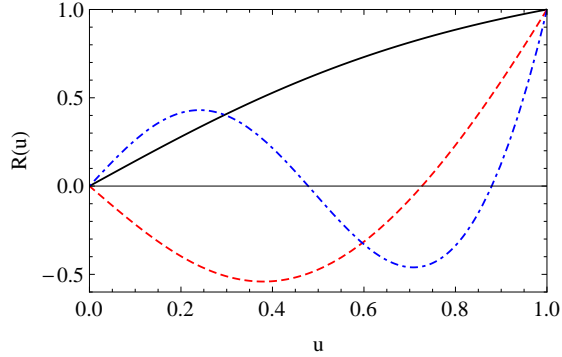}
\caption{The lowest-lying marginally stable modes for the Gauss-Bonnet parameter $\alpha=1/20$ and $\Delta_+=3/2$. The three curves correspond to $\zeta_0=31.27 $(black solid), $\zeta_1=179.14$ (red dashed) and $\zeta_2=443.41$ (blue dot-dashed), respectively.}
\label{figure1}
\end{figure}

\begin{figure}
\centering
\includegraphics[width=2.9in]{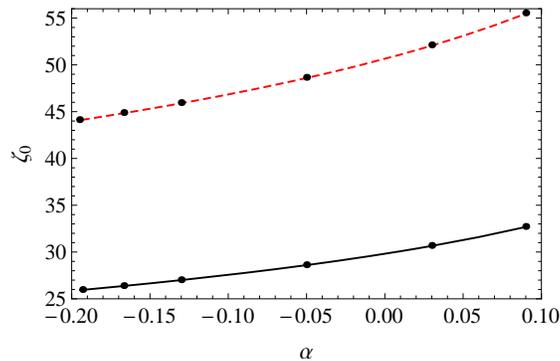}
\caption{The critical value of $\zeta=|\gamma q B|/T^2$ with respect to the Gauss-Bonnet parameter $\alpha$ with $\Delta_+=3/2$ (black solid) and $2$ (red dashed) for $E^L_0=-|q\gamma B|$. The curves are from the shooting method while the black points are obtained from the SL method.}
\label{figure2}
\end{figure}

Solving Eq.~(\ref{eomzeta}), we exhibit the first three lowest-lying marginally stable modes in Fig.~\ref{figure1} for $\alpha=1/20$ and $\Delta_+=3/2$ in Fig.~\ref{figure1}.
In these three modes, the solid curve with $\zeta_0=31.27$ does not intersect with the horizontal axis except at $u=0$, we denote this mode with the node $n=0$. Therefore, the node of other two curves is $n=1~(\zeta_1=179.14)$ and $n=2~ (\zeta_2=443.41)$, respectively. It is known to all that the node implies the oscillation of $R(u)$ along the $u$ direction, which will cost the energy of the system and thus be instable. Hence, the value $\zeta_0$ with the zero node is thought as the critical value for the superconductor phase transition. We plot the critical value $\zeta_0$ as a function of $\alpha$ with fixed $\Delta_+=\frac{3}{2}$ and $2$ in Fig.~\ref{figure2}.

As we know, for a given magnetic field $B$, when the temperature $T$ decreases to a critical value, corresponding to increasing $\zeta=|\gamma q B|/T^{2}$ in Fig.~\ref{figure2}, the normal phase will become instable to  transforming into the condensed phase. Therefore the upper left region for each boundary curve in the figure represents the condensed phase. From the figure, we find even without the charge density, the superconductor phase transition can still be induced by the applied magnetic field, which is  similar to the QCD vacuum instability induced by the strong magnetic field to trigger the vector condensate~\cite{Chernodub:2010qx,Chernodub:2011mc}, but it is opposite to the ordinary superconductor where the magnetic field hinders the phase transition~\cite{Albash:2008eh,Nakano:2008xc,Zhao:2013pva}. We clearly see from the action (\ref{Lvector}) that the nonminimal coupling term leads to the superconductor phase transition. In addition, $\zeta_0$ improves with the increasing $\alpha$. More precisely, for the given magnetic field $|\gamma q B|$, the larger the gravitational parameter $\alpha$, the lower the critical temperature, which indicates that the increasing $\alpha$ inhibits the phase transition. Moreover, for the given $\alpha$ and $|\gamma q B|$, the larger dimension of the operator $\Delta_+$ decreases the critical temperature, which means that the increasing $\Delta_+$ makes the phase transition more difficult. This is obvious from the effective mass of the vector field in Eq.~(\ref{eigvarpx}), where the larger $\Delta_+$ corresponds to the larger mass squared $m^2$; it then makes the effective mass more difficult to fall below the BF bound.

To uphold above numerical results, we then solve Eq.~(\ref{BHMCVpsiu}) by the analytical SL eigenvalue method~\cite{Siopsis}. By introducing a trial function $\Gamma(u)$ as $\psi_x(u)=\langle J_x\rangle u^{\Delta_{+}}\Gamma(u)$, we can obtain  the equation of motion for $\Gamma(u)$
\begin{eqnarray}
\Gamma ''+ \left(\frac{f'}{f}+\frac{2 \Delta
   _+-1}{u}\right)\Gamma '+\frac{\Delta _+ u f'+\left(\Delta _+-2\right) \Delta
   _+ f- m^2}{f u^2}\Gamma+ \zeta\frac{ u^2}{\pi ^2 f}\Gamma =0,
\end{eqnarray}
with the boundary conditions $\Gamma(0)=1$ and $\Gamma'(0)=0$. Such an equation is further written as the SL eigenvalue equation
\begin{equation}
\frac{d}{du}\big(\underbrace{ u^{2 \Delta_+-1}f}_{K}\Gamma '\big)+\underbrace{u^{2 \Delta _+-3} \left(\Delta _+ u   f'+\left(\Delta _+-2\right) \Delta _+ f-m^2\right)}_{-P}\Gamma +\zeta  \underbrace{\frac{ u^{2 \Delta _+-1}}{\pi^2}}_Q\Gamma=0.
\end{equation}
Hence, the critical value of $\zeta$ can be obtained by minizing the following function
\begin{equation}\label{integ}
\zeta=\frac{\int^1_0du(K{\Gamma'}^2+P\Gamma^2)}{\int^1_0duQ\Gamma^2}.
\end{equation}

Concretely, we take the trial function $\Gamma(u,a)=1-a u^2$ with the constant $a$ to be determined. Then we can obtain the minimum value of $\zeta$ from Eq.~(\ref{integ}) for a given $\alpha$. We obtain the analytical results of $\zeta_0$ for some special Gauss-Bonnet parameter $\alpha$, and plot the results in Fig.~\ref{figure2} in the form of black points, from which we find that the analytical results agree with the numerical results. Therefore, we conclude that the analytical method is powerful for this MCV model.

\begin{figure}
 \centering
  \includegraphics[width=3.2 in]{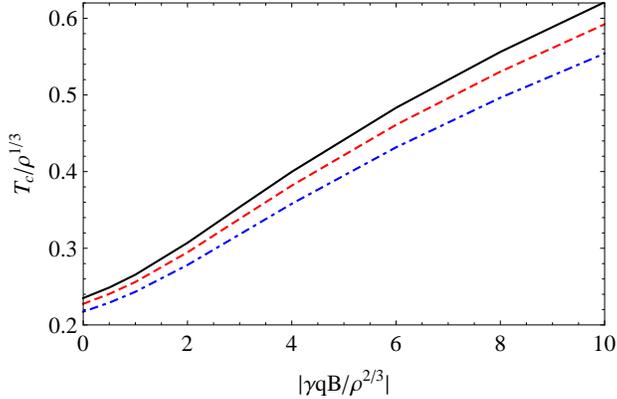}
  \caption{The critical temperature versus the magnetic field with different $\alpha$ and the fixed $\Delta_+=3/2$ for the lowest Landau level. The curves from top to bottom correspond to $\alpha=-19/100$~(black solid), $-1/20$~(red dashed), $9/100$~(blue  dot-dashed), respectively.}
\label{figure3}
\end{figure}

Next we consider such a $p$-wave superconductor model with a finite charge density $\rho$. In the presence of the electric field $\phi$, Eq.~(\ref{BHMCVpsiu}) reads
\begin{equation}
R''+ \left(\frac{f'}{f}+\frac{2 \Delta _--1}{u}\right)R'+ \left(\frac{\lambda ^2\left(u^2-1\right)^2}{f^2}-\frac{E_l \lambda ^{2/3}}{f \rho ^{2/3}}+\frac{\Delta _- u f'+\Delta _-
   f\left(\Delta _--2\right)-m^2}{f u^2}\right)R=0,
\end{equation}
where we have used the relation $\lambda=\frac{\rho}{r_+^3}$. It is clear that such equation depends on two dimensionless parameters, i.e., $E_l/\rho ^{2/3}$ and $\lambda$. Solving this equation with the regular boundary condition at the horizon and the source-free condition at infinity, we plot the critical temperature $T_c/\rho^{1/3}$ as a function of the applied magnetic field $|\gamma q B/\rho^{2/3}|$ with $\Delta_+=3/2$ for $\alpha=-19/100$~(black solid),~$-1/20$~(red dashed) and $9/100$~(blue dot-dashed) in Fig.~\ref{figure3}.
Evidently, the lower right region denotes the superconducting phase while the upper left part represents the normal phase. From the figure, we have the following comments: the strong magnetic field can still trigger the holographic superconductor phase transition, which is similar to the case without the charge density.  Since it is very reminiscent of the QCD vacuum instability triggered by the magnetic field to developing the $\rho$ meson condensate~\cite{Chernodub:2010qx,Chernodub:2011mc}, we can therefore regard our model as the holographic realization of the QCD vacuum instability to some extent. Moreover, for the fixed magnetic field, the critical temperature decreases with the increase of the Gauss-Bonnet parameter $\alpha$, which implies that the larger parameter $\alpha$ inhibits the transition, similar to the results for the ordinary superconductor~\cite{Cai:2010cv,Li:2011xja}. In particular, in the case of $B=0$, the critical behavior is triggered by the electric field. In addition, we find that when we increase the parameter $\alpha$, the response of the critical temperature to the magnetic field becomes less obvious. We also calculate some other cases of the dimension of operator, for example, $\Delta_+=2$, where the results are similar to the case of $\Delta_+=3/2$.

\begin{figure}
 \centering
  \includegraphics[width=3.2 in]{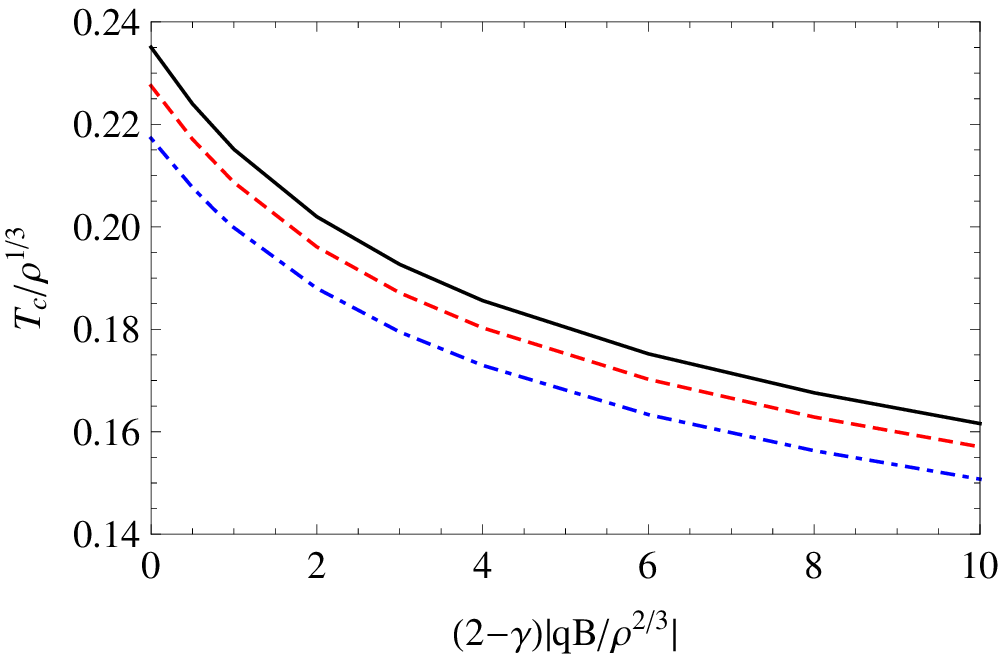}
  \caption{The critical temperature as a function of the magnetic field with different $\alpha$ and the fixed $\Delta_+=3/2$ for the non-lowest Landau level ($E_1=(2-\gamma)|q B|$). The curves from top to bottom correspond to $\alpha=-19/100$~(black solid), $-1/20$~(red dashed), $9/100$~(blue dot-dashed), respectively.}
\label{figure4}
\end{figure}

Besides, we also  plot the critical temperature $T_c/\rho^{1/3}$ as a function of $(2-\gamma)|q B/\rho^{2/3}|$ for the excited Landau level, for example, $E_1=(2-\gamma)|q B|$ (by choosing $l=1$, sign($qcB$)$=\mp$, and $\gamma>0$) in Fig.~\ref{figure4}, to compare with the case of $E_0^L=-|\gamma q B|$  in Fig.~\ref{figure3}.
Noting that when $E_1<0$ (i.e., $\gamma>2$), the effect of the magnetic field on the phase transition is similar to the one in the case of the lowest Landau level. Hence, to qualitatively illustrate the difference between the excited Landau level with $E_l>0$ and the lowest Landau level with $E_0<0$, we have chosen $E_1>0$ (i.e., $0<\gamma<2$).
In Fig.~\ref{figure4}, the lower left region denotes the superconductor phase, while the other part represents the normal phase. Moreover, we find that with the increasing magnetic field, the system  will become instable and then the superconducting phase will be broken into the normal phase, which is similar to the ordinary superconductor~\cite{Albash:2008eh,Nakano:2008xc,Zhao:2013pva}. Moreover, the critical temperature for this MCV model decreases with the increasing Gauss-Bonnet parameter.

\subsection{Yang-Mills model}
The holographic superconductor phase transition induced by the nonAbelian magnetic field  was calculated in the black hole background~\cite{Wong:2013rda}. To reveal the relation between the MCV model and the nonAbelian model, next we study the magnetic-field-induced phase transition  in the  Gauss-Bonnet-AdS black hole background coupled to an SU(2) YM field in the probe limit. Following Ref.~\cite{Gubser2008a}, the action of the SU(2) YM gauge field reads
\begin{equation}\label{Lpwave}
  \mathcal{S}_{YM}=\frac{1}{16\pi G_5}\int dx^5\sqrt{-g}\left(-\frac{1}{4}F^a_{\mu\nu}F^{a\mu\nu}\right),
\end{equation}
where the field strength of the gauge field is defined by $F^a_{\mu\nu}=\nabla_\mu A^a_\nu-\nabla_\nu A^a_\mu+\varepsilon^{abc}A^b_\mu A^c_\nu$ with the gauge index $a=1,~2,~3$. Moreover, there are three generators $\tau^i$ for this SU(2) group with the commutation relation $[\tau^i,\tau^j]=\varepsilon^{ijk}\tau^k$ ($i,~j,~k=1,~2,~3$).  Varying the action (\ref{Lpwave}) with respect to the gauge field $A=A^a_\mu\tau^a dx^\mu$, we can write out the equation of motion as
\begin{equation}\label{Eomp}
   \nabla_\mu F^{a\mu\nu}+\varepsilon^{abc}A^b_\mu F^{c\mu\nu}=0.
\end{equation}
Comparing with the ansatz of the MCV field, we take the assumptions for the SU(2) field as
\begin{eqnarray}\label{Su2ans}
A^1_\mu dx^\mu&=&\epsilon a^1_x(r,x,y) dx+\epsilon a^1_y(r,x,y)dy,\nonumber\\
A^2_\mu dx^\mu&=&\epsilon a^2_x(r,x,y)dx+\epsilon a^2_y(r,x,y)dy,\\
A^3_\mu dx^\mu&=&\phi(r)dt+B x dy,\nonumber
\end{eqnarray}
where $\epsilon$ is a small parameter and the magnetic field $B$ is perpendicular to the $x-y$ plane.

Defining $\Psi_x(r,x,y)=a^1_x-ia^2_x$ and $\Psi_y(r,x,y)=a^1_y-ia^2_y$ and substituting the ansatz (\ref{Su2ans}) into Eq.~(\ref{Eomp}), we can read off the equations of motion
\begin{eqnarray}
\partial_x \Psi_x+(\partial_y-i B x)\Psi_y&=&0,\\
\partial_x\partial_r \Psi_x+(\partial_y-i B x)\partial_r \Psi_y&=&0,
\end{eqnarray}
\begin{eqnarray}
\partial_r^2\Psi_x+\left(\frac{3}{r}+\frac{f'}{f}\right)\partial_r \Psi_x
+\frac{1}{r^4f}\bigg(\left(\partial^2_y-2 i B x \partial_y+\frac{\phi^2}{f}-B^2x^2\right)\Psi_x-\qquad\qquad&&\nonumber\\
 (\partial_x\partial_y-i B x\partial_x+i B)\Psi_y\bigg)&=&0,\\
\partial_r^2\Psi_y+\left(\frac{3}{r}+\frac{f'}{f}\right)\partial_r \Psi_y
+\frac{1}{r^4f}\left((2 i B-\partial_x\partial_y+i  B x \partial_x )\Psi_x+\left(\partial_x^2+\frac{\phi^2}{f}\right)\Psi_y\right)&=&0.
\end{eqnarray}
 To solve above four equations, we should further separate the functions $\Psi_x$ and $\Psi_y$ as
\begin{equation}
\Psi_x(r,x,y)=\tilde{\psi}_x(r)\tilde{U}(x)e^{ipy}, \ \ \ \Psi_y(r,x,y)=\tilde{\psi}_y(r)\tilde{V}(x)e^{ipy}e^{i \theta},
\end{equation}
which further yields the equations of motion
\begin{eqnarray}
\tilde{\psi}_x (r) \dot{\tilde{U}}(x)\pm (B x-p)\tilde{\psi}_y (r) \tilde{V}(x)=0, \label{tilU}\\
\tilde{\psi}'_x (r) \dot{\tilde{U}}(x)\pm (B x-p)\tilde{\psi}'_y (r) \tilde{V}(x)=0,\label{tilV}
\end{eqnarray}
\begin{eqnarray}
\tilde{\psi}''_x+\left(\frac{3}{r}+\frac{f'}{f}\right)\tilde{\psi}'_x+\frac{\phi^2}{r^4f^2}\tilde{\psi}_x+
\frac{\tilde{\psi}_x}{r^4f}\left(-(B x-p)^2\pm\left((p-B x)\frac{\dot{\tilde{V}}}{\tilde{U}}
+\frac{B \tilde{V}}{\tilde{U}}\right)\frac{\tilde{\psi}_y}{\tilde{\psi}_x}\right)&=&0,\label{tilvax} \\
\tilde{\psi}''_y+\left(\frac{3}{r}+\frac{f'}{f}\right)\tilde{\psi}'_y+
\frac{\phi^2}{r^4f^2}\tilde{\psi}_y
+\frac{\tilde{\psi}_y}{r^4f}\left(\frac{\ddot{\tilde{V}}}{\tilde{V}}\pm \left( 2B\frac{\tilde{U}}{\tilde{V}}+ (B x-p )\frac{\dot{\tilde{U}}}{\tilde{V}}\right)\frac{\tilde{\psi}_x}{\tilde{\psi}_y}\right)&=&0,\label{tilvay}
\end{eqnarray}
where the dot and the prime stand for the derivative with respect to $x$ and $r$, respectively, while the ``$\pm$" corresponds to the phase difference $\theta_\pm=\pm\frac{\pi}{2}+2 n\pi$ with an integer $n$. By comparing Eqs.~(\ref{tilU})-(\ref{tilvay}) with Eqs.~(\ref{eomX})-(\ref{eomvarpy}), it is easy to see that  the SU(2) model is a special case of the MCV model with the parameters chosen as $m^2=0,~q=1$, and $\gamma=1$, which has been observed in Refs.~\cite{Cai:2013kaa,wuyb}. Therefore, the results for the SU(2) model can be directly obtained by taking the parameters of the  MCV model $m^2=0,~q=1$, and $\gamma=1$, and the effects of the Gauss-Bonnet parameter on the critical value are also similar to the case of the MCV model whether the Landau level is the lowest or not.
\section{$p$-wave phase transition in Gauss-Bonnet-AdS soliton }
In this section, we study the MCV model in the Gauss-Bonnet-AdS soliton background. As we know, the soliton solution can be obtained from the black hole solution via a double Wick rotation. By using this rotation ($t\rightarrow i\eta,~z\rightarrow i t$) to black hole~(\ref{Gmetric}), the Gauss-Bonnet-AdS soliton  is of the form~\cite{Cai:2007wz}
\begin{equation}\label{GBSmetric}
ds^2=-r^2dt^2+\frac{dr^2}{r^2f(r)}+r^2(dx^2+dy^2)+r^2f(r)d\eta^2,
\end{equation}
where the function $f(r)$ is the same as that in Eq.~(\ref{Gmetric}). To distinguish the soliton from the black hole, we denote the tip with $r_s$ satisfying $f(r_s)=0$. Moreover, to have a smooth geometry at the tip, we should impose a period $\eta\sim\eta+\pi/r_s$ on the spatial direction $\eta$ for the Scherk-Schwarz circle. Since the physical region of this soliton spacetime is only for $r>r_s$, the geometry is dual to a confined phase with an energy gap in the dual field theory and thus can be used to describe the insulator  in condensed matter physics. In this section, we still constraint the range of $\alpha$ as $-7L^2/36\leq\alpha\leq9L^2/100$, while the asymptotical form of $f(r)$ and the effective AdS radius still take the forms as Eqs.~(\ref{asyform}) and (\ref{Adsr}), respectively.

For the MCV model, we still turn on the ansatz as Eqs.~(\ref{MCVansatz}) and (\ref{MCVansAu}). By using complicated calculations, we can obtain three equations of motion in terms of $\psi_x(r)$ and $U(x)$ as well as $V(x)$, where the equation for $\psi_x(r)$ reads
 \begin{equation}\label{SMCVpsixr}
\psi''_x+\left(\frac{3}{r}+\frac{f'}{f}\right)\psi'_x-\frac{m^2}{r^2f}\psi_x+\frac{q^2\phi^2}{r^4f}\psi_x- \frac{E}{r^4f}\psi_x=0,
\end{equation}
where the prime denotes the derivative with respect to $r$, and the constant $E$ is the eigenvalue, while the equations for $U(x)$ and $V(x)$ are exactly the same as Eqs.~(\ref{eigX}) and (\ref{eigY}), respectively, and hence the Landau level is still of the form (\ref{Enphi}).

Substituting the gauge field ansatzs (\ref{MCVansAu}) and (\ref{MCVansatz}) into Eq.~(\ref{EOMphi}), at the linear order, the equation in terms of $\phi(r)$ in the background (\ref{GBSmetric}) reads
\begin{equation}
\phi''+(\frac{3}{r}+\frac{f'}{f})\phi'=0.
\end{equation}
From the gauge/gravity dual dictionary, the leading term of the general expansion of the gauge field near the boundary $r\rightarrow \infty$ is dual to the chemical potential in the boundary field theory. Considering further the Neumann-like boundary conditions at the tip~\cite{Nishioka:2009zj}, we can obtain the solution of $\phi(r)$ as
\begin{equation}
\phi(r)=\mu.
\end{equation}

We now study the effects of the applied magnetic field and the Gauss-Bonnet parameter $\alpha$ on the critical point between the insulator phase and the superconducting phase. To calculate the infinite boundary behavior of the vector field, it is convenient to introduce a new coordinate $u=r_s/r$, by using which we can rewrite Eq.~(\ref{SMCVpsixr}) as
\begin{equation}\label{SMCVpsiu}
\psi _x''(u)+\left(\frac{f'(u)}{f(u)}-\frac{1}{u}\right) \psi _x'(u)+ \left(\frac{\zeta}{f(u)}-\frac{m^2}{u^2 f(u)}\right)\psi _x(u)=0,
\end{equation}
with a dimensionless parameter $\zeta=\frac{q^2 \mu^2-E_l}{r_s^2}$ determining the critical behavior of the system. Near the boundary $u\rightarrow0$, the general expansion of $\psi_x(u)$ is the same as Eq.~(\ref{expvarx}) with $\Delta_\pm=1\pm\sqrt{1+ m^2 L_{eff}^2}$. For the purpose of numerical calculation, defining a new function $\psi_x (u)=u^{\Delta_-}R(u)$ results in
\begin{equation}
R''+ \left(\frac{f'}{f}+\frac{2 \Delta _--1}{u}\right)R'+\left(\frac{\Delta _- f'}{uf}+\frac{\Delta _-(\Delta _--2)}{u^2}-\frac{m^2}{u^2f}+\frac{\zeta}{f}\right)R =0.
\end{equation}

To solve this equation via the shooting method, we usually impose the regular boundary condition at the tip, for example, $R(1)=1$, as well as the source-free boundary condition at the boundary $u\rightarrow 0$, which requires the leading term of the general falloff of $R(u)$ to be zero. Since we are interested in the critical behavior, as discussed in the previous section, for a given  $m^2$ and $\alpha$, only a special $\zeta$ can give the lowest-lying marginally stable mode and thus stands for the critical value where the phase transition takes place.

\begin{figure}
 \centering
  \includegraphics[width=3.2 in]{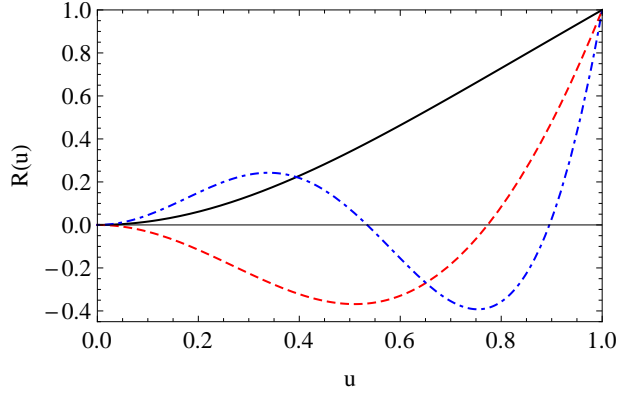}\\
  \caption{The lowest-lying marginally stable modes of the vector field with $\alpha=1/20$ and $\Delta_+=2$ for the lowest Landau level ($E_0^L=-|\gamma q B|$). The curves correspond to $\zeta_0=5.38$~(black solid), $\zeta_1=23.36$~(red dashed), $\zeta_2=53.12$~(blue dot-dashed), respectively.}
  \label{figure5}
\end{figure}

\begin{figure}
 \centering
  \includegraphics[width=3.2 in]{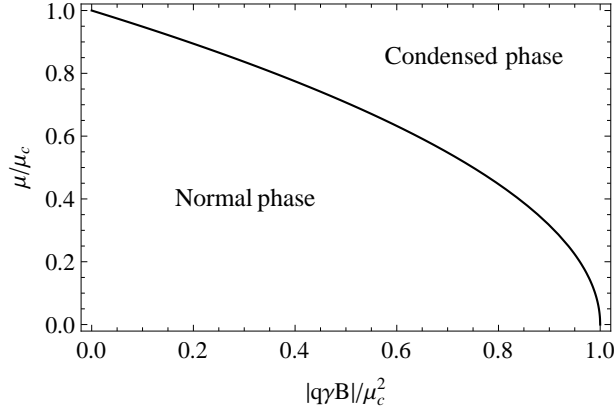}\\
  \caption{The critical chemical potential as a function of the magnetic field for the lowest Landau level ($E_0^L=-|\gamma q B|$), where $\mu_c=\sqrt{r^2_s\zeta_0}$.}
  \label{figure6}
\end{figure}

We consider the lowest Landau level, i.e., $E_0^L=-|\gamma q B|$. In the case of $\alpha=1/20$ and $\Delta_+=2$, we show the first three lowest-lying marginally stable modes in Fig.~\ref{figure5}.
Since the mode with the nonzero node implies the instability of the system, we take the mode with the zero node ($\zeta_0=5.38$) as the critical value where the vector condensate emerges. From this approach, we calculate the critical chemical potential $(\mu/\mu_c)$ versus the magnetic field $(|q \gamma B|/{\mu_c^2})$ in Fig.~\ref{figure6}.
 Clearly, as we increase the applied magnetic field to a critical value,  the insulator/superconductor phase transition is triggered, which is similar to the QCD vacuum  instability induced by the strong applied magnetic field~\cite{Chernodub:2010qx,Chernodub:2011mc} but opposite from the ordinary critical behavior~\cite{Cai:2011tm}. Moreover, the critical chemical depends on the Gauss-Bonnet parameter $\alpha$ and the mass squared of the vector field $m^2$, while Fig.~\ref{figure6} is independent of $\alpha$ and $m^2$, which can be understood as follows: for different  $\alpha$ and $m^2$, the critical chemical potential is different, because it is calculated in the absence of the magnetic field, i.e., $\mu_c^2=r^2_s \zeta_0$. However, if we plot the chemical potential versus the magnetic field by scaling the unit of $\mu_c$, the functional relation can be written as $(\mu/\mu_c)^2+(|q \gamma B|/{\mu_c^2})=1$. Therefore, the relation between $(\mu/\mu_c)$ and $(|q \gamma B|/{\mu_c^2})$ is independent of $\alpha$ and $m^2$.

\begin{figure}
  \centering
  \includegraphics[width=3.2 in]{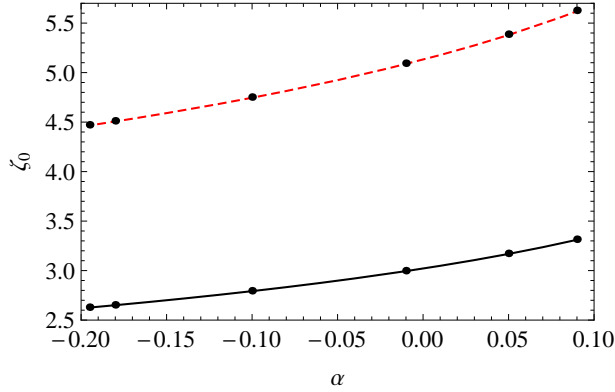}\\
  \caption{The critical value of $\zeta=(q^2 \mu^2-E_l)/r_s^2$ as a function of the Gauss-Bonnet parameter $\alpha$   for the lowest Landau level ($E_0^L=-|\gamma q B|$). The curves from the shooting method correspond to $\Delta_+=3/2$~(black solid), and 2~(red dashed), respectively, while the black points are from the SL method.}
  \label{figure7}
\end{figure}

In addition, in Fig.~\ref{figure7}, we plot the critical value of $\zeta$ corresponding to the marginally stable mode with zero node for various Gauss-Bonnet parameter $\alpha$ in the case of the lowest Landau level, where the upper curve denotes the dimension of the operator $\Delta_+=2$ and the lower curve corresponds to $\Delta_+=3/2$.
From the figure, we conclude that the critical value $\zeta_0$  improves with the increase of the Gauss-Bonnet parameter $\alpha$. If we turn off the magnetic field $B$ (or the chemical potential $\mu$), it is clear that $\mu_c$ (or $B_c$) improves with the increasing $\alpha$, which means that the larger $\alpha$ makes the phase transition more difficult. Moreover, for a given $\alpha$, the critical value $\zeta_0$ in the case of $\Delta_+=2$ is larger than that for $\Delta_+=3/2$. The effect of $B$ and $\Delta_+$ on the critical value $\zeta_0$ can be understood from the effective mass of the vector field. From Eq.~(\ref{SMCVpsixr}), it is easy to see that both the increasing $B$ and the larger $\Delta_+$ corresponding to the increasing $m^2$ improve the effective mass, and thus hinder the superconductor phase transition.

To backup the numerical results, we recalculate the critical value $\zeta_0$ by the analytical SL method. As worked in the Gauss-Bonnet-AdS black hole background, by defining a trial function $\Gamma(u)$ as $\psi_x(u)=\langle J_x\rangle u^{\Delta_{+}}\Gamma(u)$, we can construct the SL eigenvalue equation in term of $\Gamma$ from Eq.~(\ref{SMCVpsiu})
\begin{equation}
\frac{d}{du}\big(\underbrace{ u^{2 \Delta_+-1}f}_{K}\Gamma '\big)+\underbrace{u^{2 \Delta _+-3} \left(\Delta _+ u   f'+\left(\Delta _+-2\right) \Delta _+ f-m^2\right)}_{-P}\Gamma +\zeta  \underbrace{u^{2 \Delta _+-1}}_Q\Gamma=0.
\end{equation}
with the boundary conditions $\Gamma(0)=1$ and $\Gamma'(0)=0$. According to Eq.~(\ref{integ}), we find the critical value $\zeta_0$ as a function of the Gauss-Bonnet parameter $\alpha$ for $\Delta_+=3/2$ and~$2$. The values of $\zeta_0$ with respect to $\alpha$ are plotted in Fig.~\ref{figure7} in the form of black points, from which we clearly see that the analytical results agree with the numerical results.

\begin{figure}
\centering
  \includegraphics[width=3.2 in]{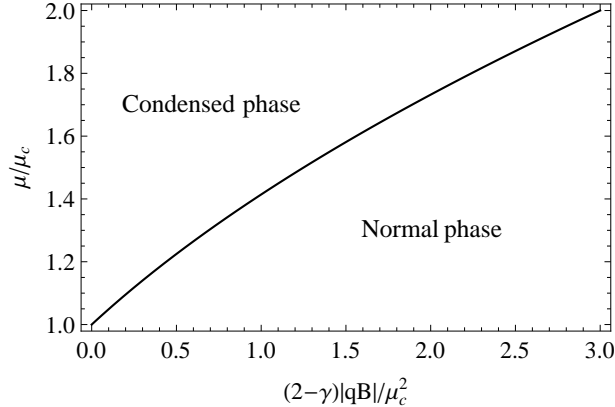}\\
  \caption{The critical chemical potential as a function of the magnetic field for the excited Landau level ($E_1=(2- \gamma)|q B|$), where $\mu_c=\sqrt{r^2_s\zeta_0}$.}
  \label{figure8}
\end{figure}

Apart from the case of the lowest Landau level, by using the shooting method, we also show the critical chemical potential $(\mu/\mu_c)$ with respect to  the magnetic field $((2- \gamma)|q B|/{\mu_c^2})$ in the case with the Landau level $E_1=(2- \gamma)|q B|$ in Fig.~\ref{figure8} (where we have considered $l=1$, sign($qcB$)$=\mp$, and $0<\gamma<2$),
from which we see the increasing magnetic field will break the superconducting phase and tend to make the normal phase more stable, which is similar to the ordinary superconductor model~\cite{Cai:2011tm,Cai:2013pda}. What is more, Fig.~\ref{figure8} has nothing to do with the Gauss-Bonnet parameter $\alpha$  and the mass squared of the vector field $m^2$, which is obvious due to the scaling unit $\mu_c$.

Similar to the calculation in Sec.~II, we also calculate the critical behavior induced by the applied magnetic field in the Gauss-Bonnet-AdS soliton background coupled to the SU(2) gauge field, with the ansatz of the gauge field same as Eq.~(\ref{Su2ans}). By using complicated calculations, we find the SU(2) YM model is still a special case of the MCV model with the parameters chosen as $m^2=0,~\gamma=1$, and $q=1$. Therefore, the results of the SU(2) model are qualitatively the same as the case of the  MCV model.

\section{Conclusions And Discussions}
So far, working with the probe approximation, we have studied the holographic $p$-wave superconductor phase transition in the Gauss-Bonnet gravity via numerical and analytical methods. Concretely, we mainly studied the effects of the applied magnetic field on the MCV model in the five-dimensional Gauss-Bonnet-AdS black hole and  soliton backgrounds, respectively. The main conclusions can be summarized as follows.

In the five-dimensional Gauss-Bonnet-AdS black hole, even without the charge density, the superconductor phase transition can still be induced by the applied magnetic field. Essentially, it is the nonminimal coupling term between the vector field $\rho_\mu$ and the U(1) gauge field that triggers the vector condensate. In the case of the lowest Landau level $E_0^L=-|\gamma qB|$, the increasing magnetic field enhances the emergence of the vector ``hair", which is rather similar to the QCD vacuum instability induced by the strong magnetic field~\cite{Chernodub:2010qx,Chernodub:2011mc}, while the stronger magnetic field makes the system more difficult to develop the vector condensate for the excited Landau level, which is reminiscent of the ordinary superconductor~\cite{Albash:2008eh,Nakano:2008xc,Zhao:2013pva}. All these appearances perhaps result from the diamagnetic and Pauli pair breaking effect of the magnetic field.  Moreover, with the increase of the Gauss-Bonnet parameter $\alpha$, the critical temperature always decreases regardless of the charge density and whether the system is at the lowest Landau level, which implies that the increasing $\alpha$ always inhibits the superconductor phase transition. In addition, we find the analytical results are in agreement with the numerical results, and the MCV field model is always a generalization of the SU(2) YM model with the general mass squared $m^2$, the charge $q$ and the gyromagnetic ratio $\gamma$. The results in the five-dimensional Gauss-Bonnet-AdS soliton are similar to the case in the Gauss-Bonnet-AdS black hole background, while the difference is that, in the AdS soliton, we obtained the curve for the critical chemical potential $\mu/{\mu_c}$ as a function of the external magnetic field $|q\gamma B|/{\mu_c^2}$ for the lowest Landau level, which is independent of $\alpha$ and $m^2$ due to the fact that the scaling unit $\mu_c$ is calculated by $\mu_c^2=r_s^2 \zeta_0$.

In a word, the results showed that the magnetic field enhances the holographic $p$-wave superconductor phase transition with the lowest Landau level, which is similar to the QCD vacuum instability, and the increasing Gauss-Bonnet parameter $\alpha$ always makes the vector condensate more difficult. Moreover, it is universal that the MCV model is a  generalization of the SU(2) YM model.  Furthermore, related studies such as Refs.~\cite{Levy,Uji} suggested that the magnetic field can indeed trigger the superconductor phase transition. Therefore, our results shed light on understanding the strong interacting system from the perspective of the gravity/gauge duality to some extent. It is worth stressing that we study the MCV model and the  SU(2) model at the probe approximation. To further understand the process of the holographic superconductor phase transition, it is our task in the near future to study the backreaction of this MCV model without the external magnetic field  in the Gauss-Bonnet-AdS gravity and then to find the boundaries of the phase diagram in the parameter spacetime.
\acknowledgments
We would like to thank Prof. R.~G. Cai and Dr. L.~Li for their helpful discussions and comments. This work is supported by the National Natural Science Foundation of China (Grant No. 11175077), the Natural Science Foundation of Liaoning Province (Grant No. L2011189) and the Ph.D Programs of Ministry of China (Grant No. 20122136110002).

\end{document}